\documentclass[aps,showpacs,preprintnumbers,amsmath,amssymb,nofootinbib]{revtex4}
\usepackage{amssymb}
\usepackage{amsfonts}
\usepackage{amsmath}
\usepackage{graphicx}
\usepackage{dcolumn}
\usepackage{bm}
\usepackage[latin1]{inputenc}
\usepackage[dvips]{hyperref}

\usepackage{graphicx}
\def\be{\begin{equation}}
 \def\ee{\end{equation}}
 \def\bea{\begin{eqnarray}}
 \def\eea{\end{eqnarray}}

\newcommand{\fr}{\frac}

\def\PR#1{{Phys.\ Rev.\ D \bf #1}}
\def\PRL#1{{Phys.\ Rev.\ Lett.\ \bf #1}}

\def\a{\alpha}

\begin{document}

\title{Four-Dimensional Asymptotically AdS Black Holes with Scalar Hair}
\author{P. A. Gonz\'{a}lez}
\email{\,\,\,\,pablo.gonzalez@udp.cl} \affiliation{Facultad de
Ingenier\'{i}a, Universidad Diego Portales, Avenida Ej\'{e}rcito
Libertador 441, Casilla 298-V, Santiago, Chile.}
\author{Eleftherios Papantonopoulos}
\email{\,\,\,\,lpapa@central.ntua.gr} \affiliation{Department of
Physics, National Technical University of Athens, Zografou Campus
GR 157 73, Athens, Greece.}
\author{Joel Saavedra}
\email{\,\,\,\,joel.saavedra@ucv.cl} \affiliation{Instituto de
F\'{i}sica, Pontificia Universidad Cat\'olica de Valpara\'{i}so,
Casilla 4950, Valpara\'{i}so, Chile.}
\author{Yerko V\'{a}squez}
\email{\,\,\,\,yerko.vasquez@ufrontera.cl}
\affiliation{Departamento de Ciencias F\'{i}sicas, Facultad de
Ingenier\'{i}a, Ciencias y Administración, Universidad de La
Frontera, Avenida Francisco Salazar 01145, Casilla 54-D, Temuco,
Chile.}
\affiliation{Departamento de F\'{\i}sica, Facultad de Ciencias, Universidad de La Serena,\\ 
Avenida Cisternas 1200, La Serena, Chile.}
\date{\today}

\begin{abstract}

We present a new family of asymptotically AdS four-dimensional
black hole solutions  with scalar hair of a gravitating system
consisting of a scalar field minimally coupled to gravity with a
self-interacting potential. For a certain profile of the scalar
field we solve the Einstein equations and we determine the scalar
potential.  Thermodynamically we show that there is a critical
temperature below which there is  a phase transition
 of a black hole with hyperbolic horizon to the new  hairy
black hole configuration.

\end{abstract}

\maketitle


\section{introduction}

Hairy black holes are interesting solutions of Einstein's Theory
of Gravity and also of certain types of Modified Gravity Theories.
These solutions
 have been extensively studied over the years mainly in connection
 with the no-hair theorems. The recent developments in string theory and
specially the application of the AdS/CFT principle to condense
matter phenomena like superconductivity (for a review see
\cite{Hartnoll:2009sz}), triggered the interest of further  study
of the behaviour of matter fields outside the black hole horizon
\cite{Gubser:2005ih,Gubser:2008px}. There are also  very
interesting recent developments in Observational Astronomy. High
precision astronomical observations of the supermassive black
holes may pave the way to  experimentally test the no-hair
conjecture
 \cite{Sadeghian:2011ub}. Also there are numerical investigations
 of single and binary black holes in the presence of scalar fields
\cite{Berti:2013gfa}.

These developments put forward the necessity  of a better
understanding of the behaviour of matter fields all the way from
the black hole horizon to asymptotic infinity. The basic physical
requirement of a black hole solution with scalar hair is the
scalar field to be regular on the horizon and to fall off
sufficiently fast at the infinity. Another important problem which
is still open is to find a way to characterize the presence of the
scalar hair. All the hairy black hole solutions known so far are
characterized by secondary hair i.e. parameters which are
connected in some way with the physical parameters of the
solution. This implies that it is not possible to continuously
connect the hairy configuration with mass $M$ and a configuration
with the same mass and no scalar field. It is desirable to find
hairy black hole solutions in which the scalar hair is
characterized by a primary hair, i.e. a conserved charge. This may
be archived if we could find a stealth solution in which the
scalar sector of the theory is hidden from the gravity sector.
Work on this direction is in progress from various groups.

The first attempts to couple a scalar field to gravity was done in
a asymptotically flat spacetime. Then hairy black hole solutions
were found \cite{BBMB} but soon it was realized that these
solutions were not examples of hairy black hole configurations
violating the no-hair theorems because they were not physically
acceptable as the scalar field was divergent on the horizon and
stability analysis showed that they were unstable
\cite{bronnikov}. To remedy this a regularization procedure has to
be used to make the scalar field finite on the horizon.

The easiest way to make the scalar field regular on the horizon is
to introduce a scale in the gravity sector of the theory through a
cosmological constant. The resulting black hole solutions with the
presence of the cosmological constant have regular scalar field on
the horizon and all possible infinities are hidden behind the
horizon. Hairy black hole solutions were found with a minimally
coupled scalar field and a self-interaction potential in
asymptotically dS space \cite{Zloshchastiev:2004ny} and also a
numerical solution was presented in \cite{Torii:1998ir}, but it
was unstable. If the scalar field is non-minimally coupled a hairy
black hole configuration was found \cite{Martinez:2002ru}, but
perturbation analysis showed the instability of the solution
\cite{Harper:2003wt,Dotti:2007cp}.
 In the case of a
negative cosmological constant, stable solutions were found
numerically for spherical geometries \cite{Torii:2001pg,
Winstanley:2002jt} and an exact solution in asymptotically AdS
space with hyperbolic geometry was presented in
\cite{Martinez:2004nb} and generalized later to include charge
\cite{Martinez:2005di}. In all the above solutions the scalar
field was conformally coupled to gravity. A generalization to
non-conformal solutions was discussed in \cite{Kolyvaris:2009pc}.
Further hairy solutions in the presence of a cosmological constant
were reported in
\cite{Anabalon:2012ta,Anabalon:2012ih,Anabalon:2012tu,Bardoux:2012tr}
with various properties. More recently new hairy black hole
solutions, boson stars and numerical rotating hairy black hole
solutions were reported
\cite{Dias:2011at,Stotyn:2011ns,Dias:2011tj,Kleihaus:2013tba,Buchel:2013uba}.

Another way to make the scalar field regular on the horizon is to
introduce a scale in the scalar sector. This can be done if in the
Einstein-Hilbert action there is a coupling of a scalar field to
Einstein tensor. The derivative coupling has the dimension of
length square and it was shown that acts as an effective
cosmological constant \cite{Amendola:1993uh,Sushkov:2009hk}. Then
in \cite{Kolyvaris:2011fk}  a gravitating system of vanishing
cosmological constant consisting of an electromagnetic field and a
scalar field coupled to the Einstein tensor was discussed. A
Reissner-Nordstrom black hole undergoes a second-order phase
transition  to a hairy black hole of generally anisotropic hair at
a certain critical temperature which we compute. The no-hair
theorem is evaded due to the coupling between the scalar field and
the Einstein tensor. Spherically symmetric hairy black hole
solutions with scalar hair were also found in the presence of this
coupling \cite{Kolyvaris:2013zfa}. Finally hairy black hole
configurations were found when a charged C-metric is conformally
coupled to a scalar field \cite{Charmousis:2009cm,
Anabalon:2009qt}. In these models the acceleration parameter plays
the role of the cosmological constant giving regularity to the
scalar field \footnote{Actually there is still an irregularity of
the scalar field on one point on the horizon but at least the
scalar field does not diverge on the whole horizon.}. In
\cite{Astorino:2013sfa} a charged C-metric coupled also to an
electromagnetic field was considered and hairy black hole
solutions were found.

The gauge/gravity duality is a principle which relates strongly
coupled systems to the their weak coupled gravity duals. One of
the most well studied system  in the context of gauge/gravity
duality, is the holographic superconductor. In its simplest form,
the gravity sector is a gravitating system with a cosmological
constant, a gauge field and a charged scalar field with a
potential (for a review see \cite{Horowitz:2010gk}). The dynamics
of the system defines a critical temperature above which the
system finds itself in its normal phase and the scalar field does
not have any dynamics. Below the critical temperature the system
undergoes a phase transition to a new configuration. From the
gravity side this is interpretated as the black hole to acquire
hair while from boundary conformal field theory site this is
interpretated as  a condensation of the scalar field and the
system enters a superconducting phase.

The whole dynamics of the holographic superconductor relies
heavily on how the scalar filed behaves outside the black hole
horizon and at infinity. In \cite{Gubser:2005ih,Gubser:2008px} a
mechanism was proposed to break an Abelian gauge symmetry near a
black hole horizon. The mechanism is similar to the
Landau-Ginzburg mechanism in superconductivity but it is more
geometrical. The effective mass of the scalar field is
$m^2_{eff}=m^2+q^2g^{tt}A^2_t$
 where $q$ is the charge of the scalar field
and $A_t$ the scalar potential of the gauge field. For large
values of the scalar charge the effective mass can become negative
signaling the breaking of the gauge symmetry outside the black
hole horizon \cite{Gubser:2008px}. However, for this mechanism to
work  the behaviour of the scalar field at infinity is crucial.
The scalar field at large distances should find a potential
barrier (the cosmological constant can be considered as the
constant term of the potential at large distances). This is
necessary because for large values of $q$ the electrostatic
repulsion overcomes the gravitational attraction and the matter
field goes to infinity. If it finds a potential barrier then it
bounces back and condenses outside the black hole horizon giving
in this way a hairy black hole configuration.

From the above discussion we conclude that it is important to
understand the behaviour of a hairy black hole at large distances.
 In this work  we
will develop a general formalism to generate asymptotically AdS
hairy black holes. We consider a gravitational system which for
simplicity consists only of a real scalar field conformally
coupled to gravity with a self-interacting  potential in which we
have incorporated the cosmological constant. We also assume
spherical symmetry of the system. We look for solutions  with the
following characteristics. The scalar field to be regular on the
horizon, to fall off at asymptotic infinity and the
self-interacting potential to go to the cosmological constant at
infinity. The strategy we follow is that we choose a profile of
the scalar field. Then the metric functions and the
self-interacting potential are calculated analytically.  A similar
procedure to construct asymptotic hairy black holes was considered
in five-dimensions \cite{Farakos:2009fx}.

The paper is organized as follows. In Section II we present the
general formalism and we apply it to some of the  existing exact
black hole solutions with a conformally coupled scalar field with
a self-interacting potential. In Section III we discuss a new
class of asymptotically AdS black holes with scalar hair. In
Section IV, we discuss the thermodynamics, first  by using the
Euclidean formalism we calculate the mass and the entropy of the
solutions and then we show that a phase transition occurs between
the asymptotically AdS four-dimensional black hole  with scalar
hair and a black hole with hyperbolic horizon and finally in
Section V we conclude.


\section{General Formalism}

In this section we will present a general formalism of a neutral
scalar field minimally coupled to curvature having a
self-interacting potential $ V(\phi)$ and using it we will review
the existing hairy black hole solutions. We start with the
Einstein-Hilbert action with  a negative cosmological constant
$\Lambda=-6l^{-2}/\kappa$, where $l$ is the length of the AdS
space and $\kappa=8 \pi G_N$, with $G_N$ the Newton constant. We will
incorporate the cosmological constant in the potential as $
\Lambda=V(0)$ ($V(0)<0$).

The  action is
 \begin{eqnarray} \label{action}
 S=\int d^{4}x\sqrt{-g}\left(\frac{1}{2 \kappa }R-\frac{1}{2}g^{\mu\nu}\nabla_{\mu}\phi\nabla_{\nu}\phi-V(\phi)\right)~.
 \end{eqnarray}

  The resulting Einstein equations from the above action are
 \begin{eqnarray}
 R_{\mu\nu}-\frac{1}{2}g_{\mu\nu}R=\kappa T^{(\phi)}_{\mu\nu}\label{field1}
 \end{eqnarray}
and the energy momentum tensor $T^{(\phi)}_{\mu\nu}$ for the
scalar field is
 \begin{eqnarray}
 T^{(\phi)}_{\mu\nu}=\nabla_{\mu}\phi\nabla_{\nu}\phi-
 g_{\mu\nu}[\frac{1}{2}g^{\rho\sigma}\nabla_{\rho}\phi\nabla_{\sigma}\phi+V(\phi)]\label{energymomentum}~.
 \end{eqnarray}
 If we use Eqs. (\ref{field1}) and (\ref{energymomentum}) we obtain the equivalent equation
 \begin{eqnarray}
 R_{\mu\nu}-\kappa\left(\partial_\mu\phi \partial_\nu\phi+g_{\mu\nu}V(\phi)\right)=0~. \label{einstein1}
 \end{eqnarray}
We consider the following metric ansatz
 \begin{eqnarray}
 ds^{2}=-f(r)dt^{2}+f^{-1}(r)dr^{2}+a^{2}(r)d \sigma^2 \label{metricBH}
 \end{eqnarray}
where $d \sigma ^2$ is the metric of the spatial 2-section, which
can have positive, negative or zero  curvature. In the case of the
metric of Eq. (\ref{metricBH}), if we use Eq. (\ref{einstein1}) we
find the following three independent differential equations
 \begin{eqnarray}
 f''(r)+2\frac{a'(r)}{a(r)}f'(r)+2V(\phi)=0~,\label{first}
 \end{eqnarray}
\begin{eqnarray}
\frac{a'(r)}{a(r)}f'(r)+\left(\frac{\left(a'(r)\right)^2}{a^{2}(r)}+\frac{a''(r)}{a(r)}\right)f(r)-\frac{k}{a^{2}(r)}+V(\phi)=0~,\label{second}
\end{eqnarray}
\begin{eqnarray}
f''(r)+2\frac{a'(r)}{a(r)}f'(r)+\left(4\frac{a''(r)}{a(r)}+2(\phi'(r))^{2}\right)f(r)+2V(\phi)=0~,\label{third}
\end{eqnarray}
where $k$ is the curvature of the spatial 2-section.  All the
quantities, in the above equations, have been rendered
dimensionless via the redefinitions $\sqrt{\kappa} \phi\rightarrow
\phi$, $\kappa \ell^{-2} V \rightarrow V$ and $r/l \rightarrow r$.

If we eliminate the potential $V(\phi)$ from the above equations
we obtain
\begin{eqnarray}
a''(r)+\frac{1}{2}(\phi'(r))^2 a(r)=0\label{adiff}~,
\end{eqnarray}
\begin{eqnarray}
f''(r)-2\left
(\frac{(a'(r))^2}{a^{2}(r)}+\frac{a''(r)}{a(r)}\right)f(r)+\frac{2k}{a^{2}(r)}=0~,\label{fdiff}
\end{eqnarray}
where the potential can be determined from Eq. (\ref{first}) if
the functions $a(r)$ and $f(r)$ are known.

To find exact hairy black hole solutions  the differential
equations (\ref{first})-(\ref{third}) have to be supplemented with
the Klein-Gordon equation of the scalar field which in general
coordinates reads \be \Box \phi =\frac{d V}{d \phi}~. \label{klg}
\ee There is a well known solution of the differential equations (\ref{first})-(%
\ref{third}) and (\ref{klg}), the MTZ solution
\cite{Martinez:2004nb}, with
the change of coordinates $\frac{dr^{\prime }}{r^{\prime 2}}=\frac{%
dr}{a(r)^{2}}$, MTZ metric is given by
\begin{equation}
ds^{2}=B\left( r^{\prime }\right) \left( -F\left( r^{\prime }\right) dt^{2}+%
\frac{1}{F\left( r^{\prime }\right) }dr^{\prime 2}+r^{\prime 2}d\sigma
^{2}\right)
\end{equation}
\begin{equation}
B(r^{\prime })=\frac{r^{\prime }(r^{\prime }+2G_{N}\mu )}{(r^{\prime
}+G_{N}\mu )^{2}}~,
\end{equation}
\begin{equation}
F(r^{\prime })=\frac{r^{\prime 2}}{l^{2}}-\Big{(}1+\frac{G_{N}\mu }{2}\Big{)}%
^{2}~.
\end{equation}
The scalar field is given by \be\phi=\sqrt{\frac{3}{4 \pi
G_N}}Arctanh\frac{G_N\mu}{r^{\prime}+G_N\mu}~, \ee where the
potential is found to be \be V(\phi) = \Lambda\sinh^2\sqrt{\fr{4
\pi G_N}{3}}\phi~. \ee  This is the simplest known hairy black
hole solution of a scalar field minimally coupled to the curvature
which goes to zero at infinity and it is regular on the horizon.
Another interesting feature of this solution is revealed if the
solution is transformed  to the Jordan frame via a conformal
transformation. Then it can be understood that its simplicity
relies on the fact that conformal symmetry in the scalar sector
allows the energy momentum tensor to be traceless resulting to a
simple relation of the scalar curvature $R$ to the cosmological
constant.

If one wants to abandon the conformal coupling of the scalar field
to curvature one can still solve the equations
(\ref{first})-(\ref{third}) and (\ref{klg}), but the resulting
potential from equation (\ref{first})  is more complicated than
the corresponding potential of the MTZ black hole solution. This solution has the form in the coordinates $\frac{dr^{\prime }%
}{r^{\prime 2}}=\frac{dr}{a(r)^{2}}$ \cite{Kolyvaris:2009pc}
\begin{equation}
B(r^{\prime })=\frac{r^{\prime }(r^{\prime }+2r_{0}^{\prime })}{%
(r^{\prime }+r_{0}^{\prime })^{2}}~,
\end{equation}%
with
\begin{equation}
F(r^{\prime })=\frac{r^{\prime 2}}{l^{2}}-g\frac{r_{0}}{l^{2}}r^{\prime }-1+g%
\frac{r_{0}^{\prime 2}}{l^{2}}-\left( 1-2g\frac{r_{0}^{\prime 2}}{l^{2}}%
\right) \frac{r_{0}^{\prime }}{r^{\prime }}\left( 2+\frac{r_{0}^{\prime }}{%
r^{\prime }}\right) +g\frac{r^{\prime 2}}{2l^{2}}\ln \left( 1+\frac{%
2r_{0}^{\prime }}{r^{\prime }}\right) ~.  \label{Psiconf}
\end{equation}%
The scalar field is
\be\phi=\sqrt{\frac{3}{4 \pi G_N}}Arctanh\frac{r^{\prime}_0}{r^{\prime}+r^{\prime}_0}~,  \ee
where $r^{\prime}_0$ is a constant and the potential is given by
\begin{eqnarray} V(\phi) &=&
\fr{\Lambda}{4\pi G_N}\sinh^2\sqrt{\fr{4 \pi G_N}{3}}\phi\nonumber\\
&+& \frac{g\Lambda}{6\pi G_N} \left[2 \sqrt{3 \pi G_N} \phi \cosh
\left(\sqrt{\fr{16 \pi G_N}{3}}\phi\right) - \frac{9}{8} \sinh
\left(\sqrt{\fr{16 \pi G_N}{3}}\phi\right)- \frac{1}{8} \sinh
\left(4 \sqrt{3 \pi
G_N}\phi\right)\right]~.\nonumber\\\label{potentialnew}
\end{eqnarray}
It is interesting to observe that there is a parameter $g$  in the
potential indicating the departure from conformal invariance. If
$g=0$ then we recover the MTZ black hole.

Another interesting solution of the equations
(\ref{first})-(\ref{third}) and (\ref{klg}) is provided  by the
C-metric solution \cite{Charmousis:2009cm}\footnote{ In
\cite{Charmousis:2009cm} there is also  an electromagnetic field
present in the action.}. The metric is given by \be ds^2=
\fr{u(y,x) }{A^2 (x-y)^2}\left[ F(y) dt^2-\frac{1}{F(y)}
dy^2+\fr{1}{G(x)} dx^2+G(x) dz^2 \right]~, \ee with  metric
functions  \bea  u(y,x) &=&1+\fr{2\pi\Lambda}{9\a}\left(
\frac{Am (x-y)}{1+Am(x+y)} \right) ^2~, \nonumber \\
F(y) &=& \frac{\Lambda}{3A^2} + 1 - y^2 - 2mA y^3 - m^2 A^2 y^4~,\nonumber \\
         G(x) &=&   1 - x^2 - 2mA x^3 - m^2 A^2 x^4~.
\eea The self-interacting  potential and the scalar field are
given by \bea \label{teos}
    V(\Psi) = \fr{\Lambda}{8\pi}\left[\cosh^4\left(\sqrt{\fr{4\pi}{3}}\Psi \right)
     + \fr{9\a}{2\pi\Lambda}\sinh^4\left(\sqrt{\fr{4\pi}{3}}\Psi \right)
    -1\right]~,
\eea \be \phi(y , x) =  \sqrt{\fr{3}{4\pi}}\mbox{Arctanh}\left(
\sqrt{-\frac{2\pi\Lambda}{9\alpha}}\:\frac{Am (x-y)}{1+Am(x+y)}
\right)~. \ee
 The limit of
$\Lambda\rightarrow 0$ is obtained upon letting the coupling
$\alpha\rightarrow 0$ so that $\frac{\alpha}{\Lambda}\sim
constant$ \footnote {The parameter $\alpha$ is the coupling
 of the scalar field in the conformal frame
\cite{Charmousis:2009cm}.}. This solution then reduces smoothly to
the solution found in \cite{BBMB} for zero acceleration $A=0$.
When $A\neq 0$ but $\Lambda=0$ the potential drops out giving the
solution of \cite{Dowker:1993bt} at minimal EM-scalar coupling. If
finally on the other hand $\Lambda\neq 0$, we obtain the solution
found in \cite{Martinez:2002ru}. A generalization of this solution
to Plebanski-Demianski spacetime was discussed in
\cite{Anabalon:2012ta} where accelerated, stationary and
axisymmetric exact solutions  with self-interacting scalar fields
in (A)dS were found.

Before ending this section we note that in \cite{Anabalon:2013sra}
a similar formalism was developed adding a dilaton field coupled
to an U(1) gauge field in the action (\ref{action}).

\section{A new class of four-dimensional asymptotically AdS black holes with scalar hair}

In this section we will follow another approach. Instead of
looking for an another exact black hole solution of the equations
(\ref{first})-(\ref{third}) and (\ref{klg}) we will look for
asymptotically AdS solutions with scalar hair. Such solutions were
studied in (2+1)-dimensions \cite{Henneaux:2002wm} and in
(3+1)-dimensions \cite{Henneaux:2004zi} in connection with
conserved charges.

Following the general formalism develop in Section II for a scalar
field coupled minimally to gravity, we consider a particular
profile of the scalar field. Consider the following ansatz for the
scalar field
\begin{equation}
\phi \left( r\right) =b\ln \left( 1+\frac{\nu }{r}\right) ~,
\label{field}
\end{equation}
where $b$ and $\nu $ are parameters. Then from equation
(\ref{adiff}) we can determine the metric function $a\left(
r\right) $ analytically
\begin{equation}\label{a}
a\left( r\right) =\alpha r^{\frac{1}{2}\left( 1+\sqrt{1-2b^{2}}\right)
}\left( r+\nu \right) ^{\frac{1}{2}\left( 1-\sqrt{1-2b^{2}}\right) }+\beta
\frac{r^{\frac{1}{2}\left( 1-\sqrt{1-2b^{2}}\right) }\left( r+\nu \right) ^{%
\frac{1}{2}\left( 1+\sqrt{1-2b^{2}}\right) }}{\nu \sqrt{1-2b^{2}}}~,
\end{equation}
where $\alpha $ and $\beta $ are integration constants. For
simplicity we take $\alpha=1 $ , $\beta =0$ and $b=1/\sqrt{2}$ and
consequently equation (\ref{a}) can be written as
\begin{equation}
a\left( r\right) =\sqrt{r\left( r+\nu \right) }~.\label{ametric}
\end{equation}

 We can also
determine the metric function $f(r)$ analytically using equation
(\ref{fdiff}). We find
\begin{equation}
f\left( r\right) =k+Fr\left( r+\nu \right) +\frac{G}{\nu
^{3}}\left( -\nu
\left( \nu +2r\right) +2r\left( r+\nu \right) \ln \left( \frac{r+\nu }{r}%
\right) \right) ~,\label{frsol}
\end{equation}
where $k=-1,0,1$ and $F$, $G$ are  constants being proportional to
the cosmological constant and to the mass respectively.

Using the solution (\ref{frsol})  from equation (\ref{first}) we
obtain the following family of self-interacting  potentials
\begin{equation}
V\left( r\right) =-\frac{1}{2\nu ^{3}r\left( r+\nu \right) }\left(
-6\nu G\left( 2r+\nu \right) +\left( 6r^{2}+6\nu r+\nu ^{2}\right)
\left( \nu ^{3}F+2G\ln \left( \frac{r+\nu }{r}\right) \right)
\right) ~. \label{potential}
\end{equation}

These potentials act as a  barrier to the scalar field at large
distances. Note that  the asymptotic behaviour of the the metric
function $f(r)$ and the scalar field is given by
\begin{equation}
f\left( r\right) =k-\frac{F\nu ^{2}}{4}+F\left( r+\frac{\nu }{2}\right) ^{2}-%
\frac{G}{3r}+O\left( \frac{1}{r^{2}}\right) ~,
\end{equation}
\begin{equation}
\phi \left( r\right) =\frac{1}{\sqrt{2}}\left( \frac{\nu }{r}-\frac{1}{2}%
\frac{\nu ^{2}}{r^{2}}+O\left( \frac{1}{r^{3}}\right) \right) ~.
\end{equation}
 The Klein-Gordon
equation (\ref{klg}) is giving the scalar field
 evolution  in its potential and at large distances
 it is trivially satisfied.

Solving equation (\ref{field}) for $r$
\begin{equation}
r=\frac{\nu }{e^{\sqrt{2}\phi }-1}
\end{equation} we can write the potential
 (\ref{potential}) in terms of the scalar field $\phi$
\begin{equation}
V\left( \phi \right) =-F\left( 2+\cosh \left( \sqrt{2}\phi \right) \right) +%
\frac{G}{\nu ^{3}}\left( 6\sinh \left( \sqrt{2}\phi \right)
-2\sqrt{2}\phi \left( 2+\cosh \left( \sqrt{2}\phi \right) \right)
\right) ~, \label{m2potn}
\end{equation}
which has a global maximum at $\phi =0$ where
$ V\left( 0\right) =\Lambda$  as expected and also $
F=-\frac{\Lambda }{3}=\frac{1}{l^{2}}$ \footnote{This potential
was also found in \cite{Anabalon:2012dw} following a different
approach.}.
Besides, we know that%
\begin{equation}
V^{\prime \prime }\left( \phi =0\right) =m^{2}~,
\end{equation}
where $m$ is the scalar field mass. Therefore, we obtain that the
scalar field mass is given by
\begin{equation}
m^{2}=\frac{2\Lambda }{3}=-2l^{-2}~,
\end{equation}
which satisfies the Breitenhner-Friedman bound that ensures the
perturbative stability of the AdS spacetime
\cite{Breitenlohner:1982jf}.

\begin{figure}[h]
\begin{center}
\includegraphics[width=0.35\textwidth]{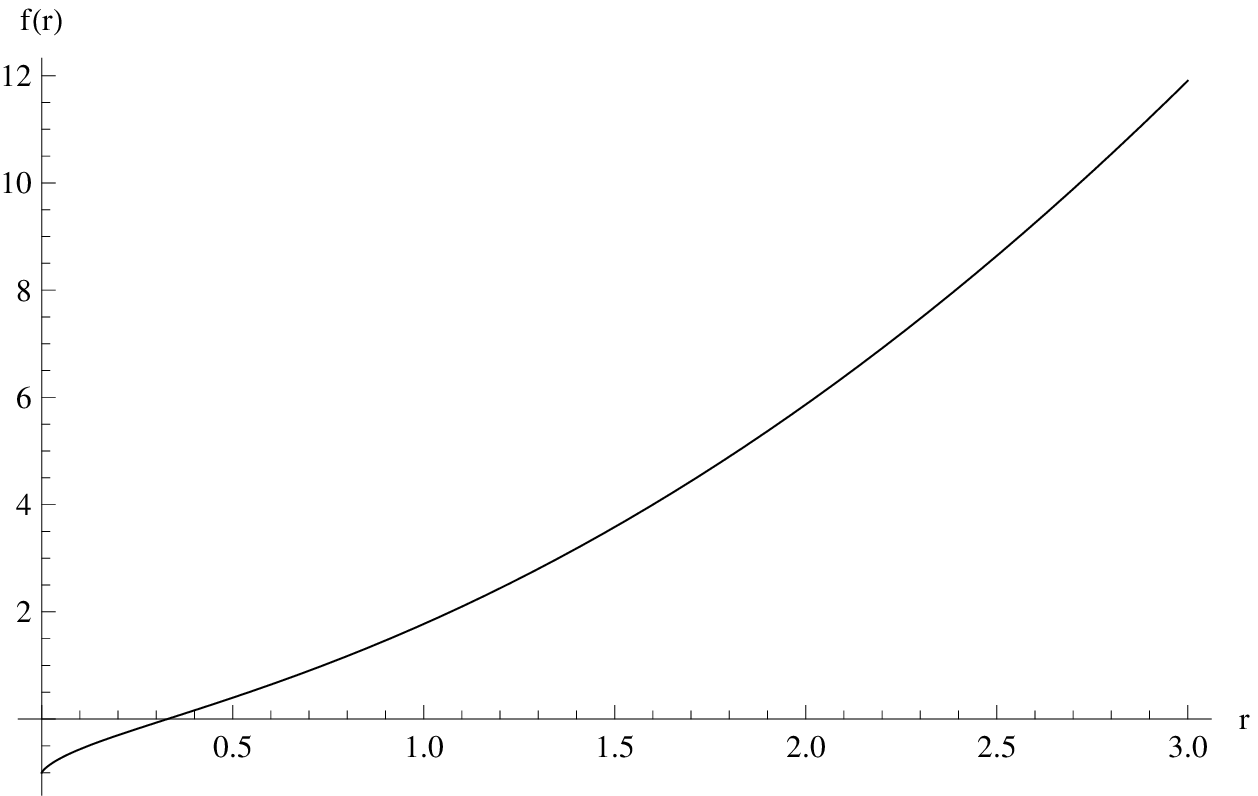}
\includegraphics[width=0.35\textwidth]{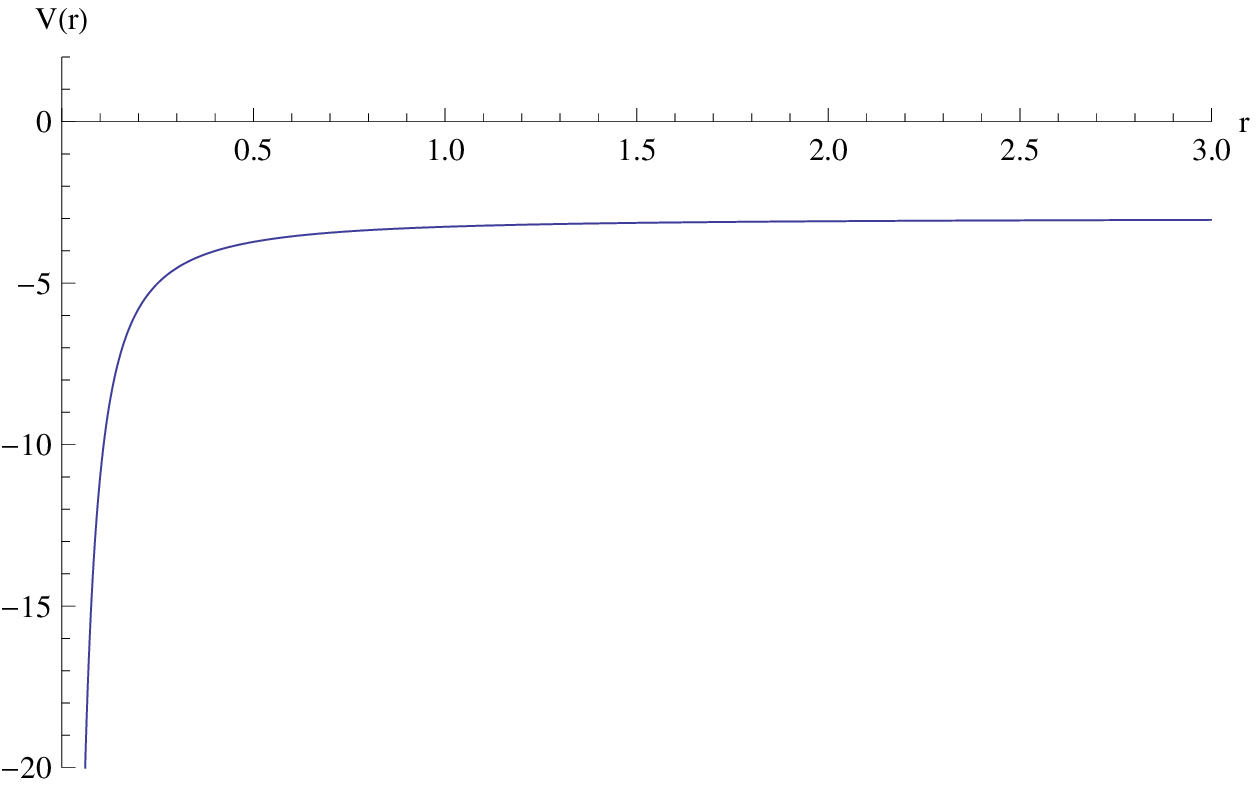}
\end{center}
\caption{The behaviour of $f(r)$ and  $V(r)$ with $k=0$, $\protect%
\nu =1$, $F=1$ and $G=1$.} \label{plots1}
\end{figure}

\begin{figure}[h]
\begin{center}
\includegraphics[width=0.3\textwidth]{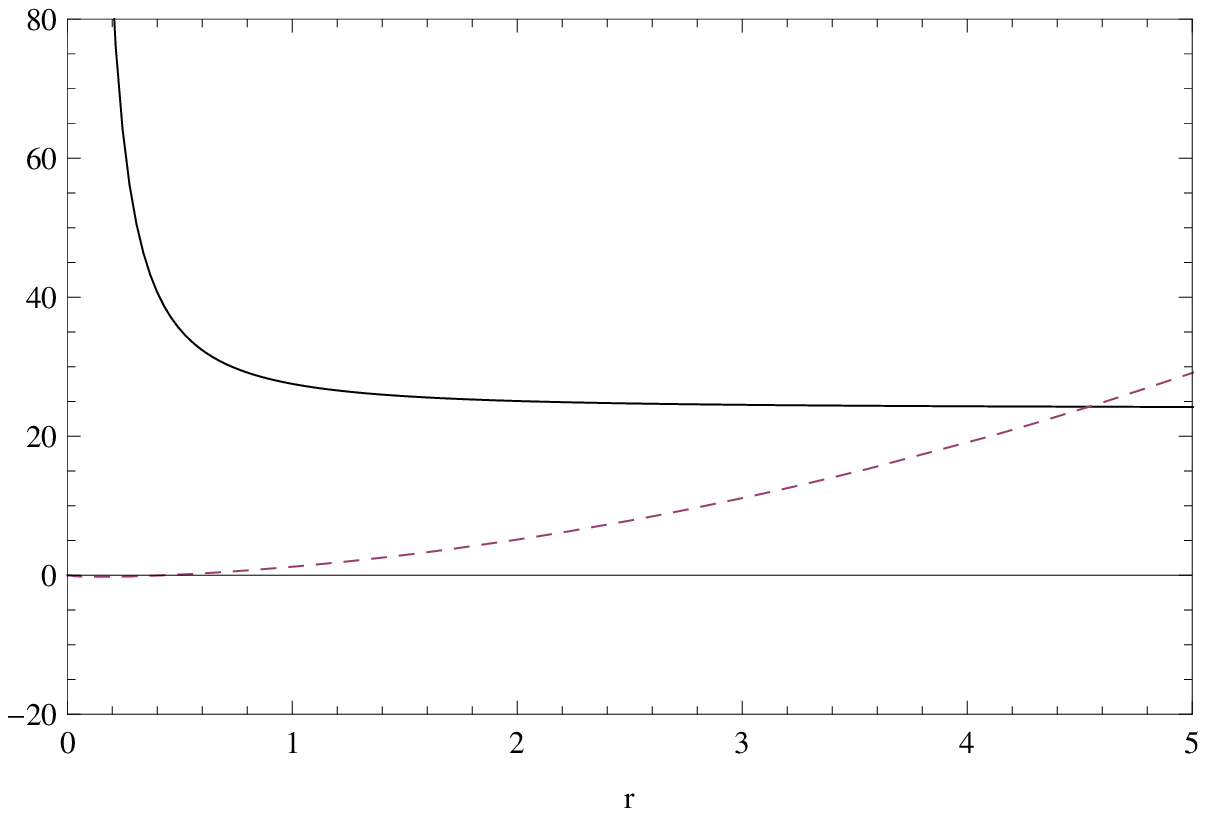}
\includegraphics[width=0.3\textwidth]{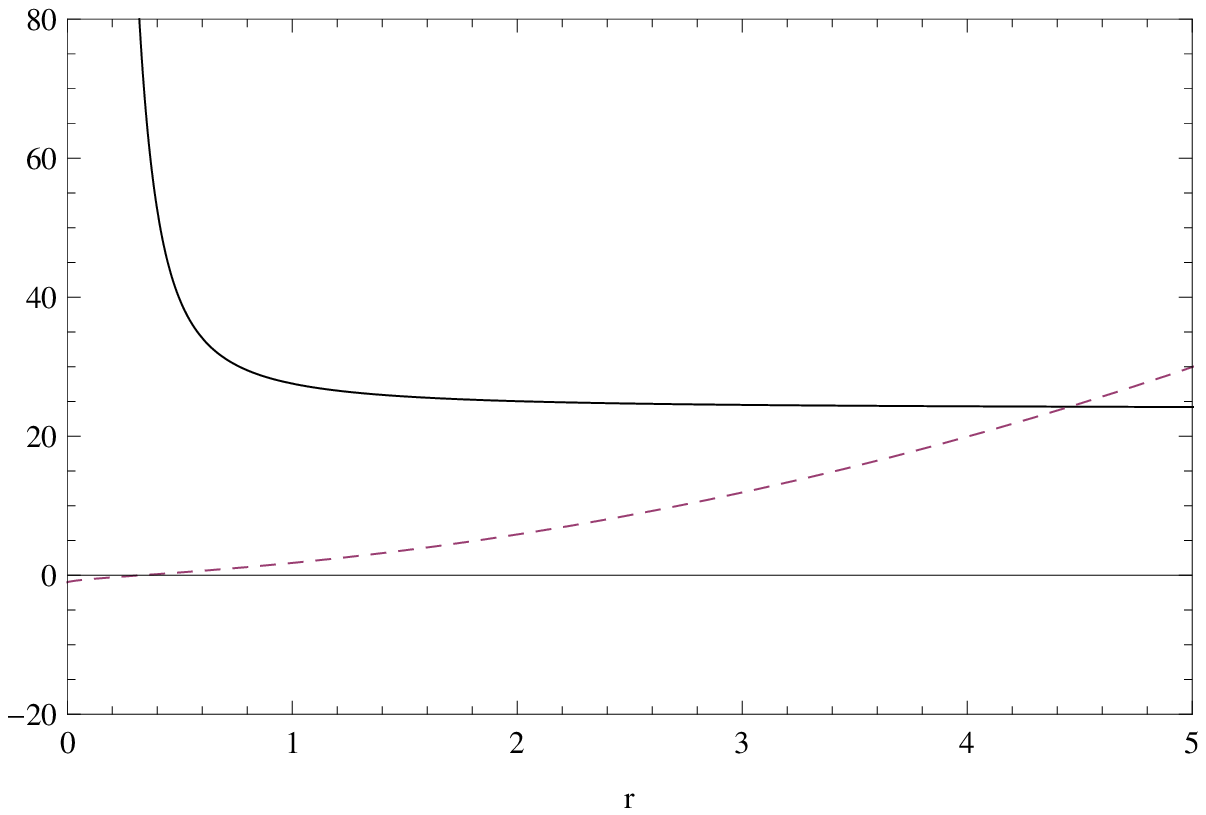}
\includegraphics[width=0.3\textwidth]{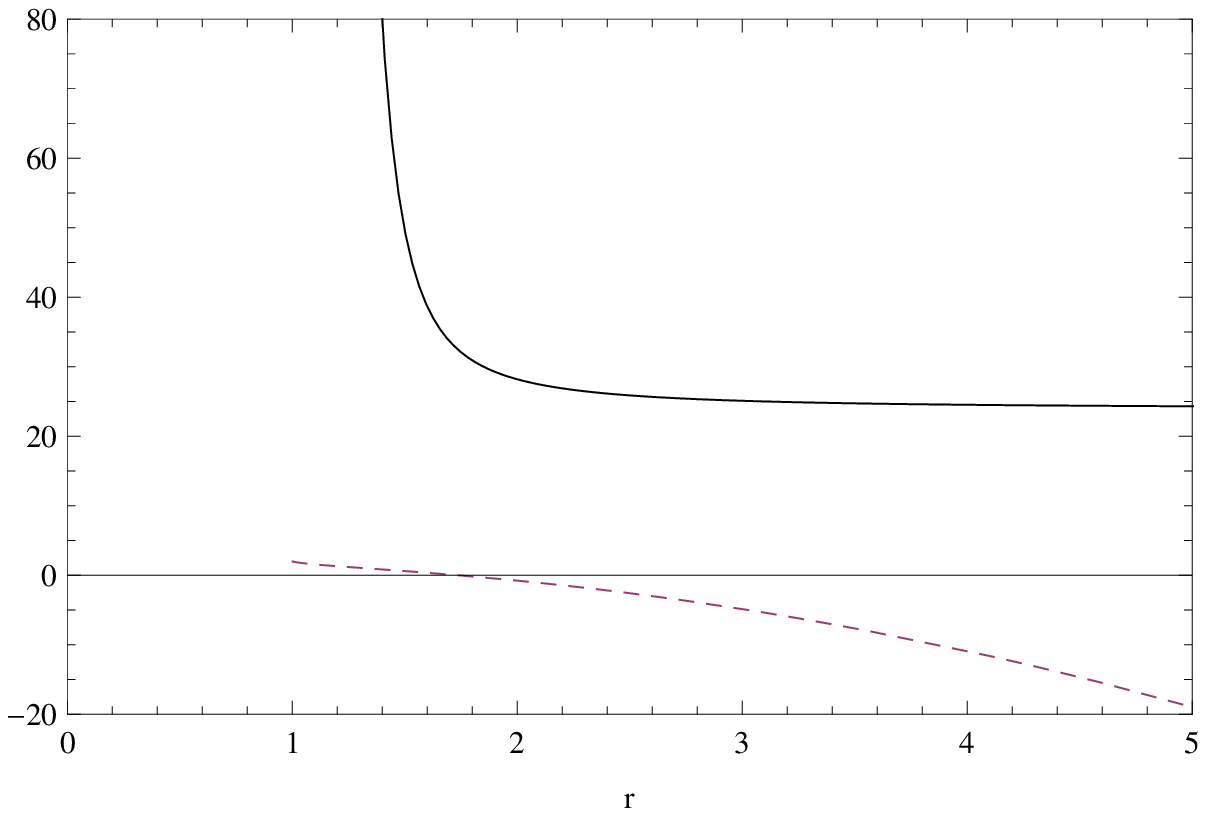}
\end{center}
\caption{The behaviour of Kretschmann scalar $R_{\mu\nu\rho\sigma}R^{\mu\nu\rho\sigma}(r)$ (Solid curve) and $f(r)$ (Dashed curve) as function of $r$. Left figure for $k=-1$, $\nu=1$, $G=-1$ and $F=1$. Center figure for $k=0$, $\nu=1$, $G=1$ and $F=1$. Right figure for $k=1$, $\nu=-1$, $G=-1$ and $F=-1$.} \label{figuraR}
\end{figure}

In Fig. \ref{plots1} we plot the  behaviour of the metric function
$f\left( r\right) $ and the potential $V\left( r\right) $ for a
choice of parameters  $k=0$, $\nu =1$, $F=1$ and $G=1$. The metric
function $f(r)$ changes sign for low values of $r$ signaling the
presence of an horizon, while the potential  asymptotically tends
to a negative constant (the cosmological constant), and  the
scalar field is regular everywhere outside the event horizon and
null at spatial infinity. We have also checked the behaviour of
the curvature outside the black hole horizon. As it is shown in
Fig. \ref{figuraR} there is no curvature singularity outside the horizon for
$k=-1,0,1$. 

To understand better the behaviour of the metric solution we have
found,  when the scalar field goes to zero asymptotically, the
metric solution should go  to the Schwarzschild anti-de Sitter
solution\footnote{This happens for $k=1$. For $k=-1$ it goes to
the topological anti-de Sitter black hole solution.}. To show that
we make a change of coordinates $\rho =\sqrt{r\left( r+\nu \right)
}$. Then the metric can be written as
\begin{equation}
ds^{2}=-\chi \left( \rho \right) dt^{2}+\frac{4\rho ^{2}/\nu
^{2}}{4\rho ^{2}/\nu ^{2}+1}\frac{1}{\chi \left( \rho \right)
}d\rho ^{2}+\rho ^{2}d\sigma ^{2}~, \label{metricnew}
\end{equation}
where
\begin{equation}
\chi \left( \rho \right) =k+F\rho ^{2}-\frac{G}{\nu }\left( \sqrt{\frac{%
4\rho ^{2}}{\nu ^{2}}+1}-2\frac{\rho ^{2}}{\nu ^{2}}\ln \left( \frac{1+\sqrt{%
\frac{4\rho ^{2}}{\nu ^{2}}+1}}{-1+\sqrt{\frac{4\rho ^{2}}{\nu ^{2}}+1}}%
\right) \right) ~.
\end{equation}
The scalar field in the new coordinates reads
\begin{equation}
\phi \left( \rho \right) =\frac{1}{\sqrt{2}}\ln \left( \frac{1+\sqrt{\frac{%
4\rho ^{2}}{\nu ^{2}}+1}}{-1+\sqrt{\frac{4\rho ^{2}}{\nu
^{2}}+1}}\right)~.\label{scafiel}
\end{equation}
At  infinity the scalar field decouples and the metric goes to
\begin{equation}
\chi \left( \rho \right) =k+F\rho ^{2}-\frac{G}{3\rho }+O\left( \frac{1}{%
\rho ^{3}}\right)~ .
\end{equation}
From the above relation we can see that the asymptotic behaviour
is the Schwarzschild anti-de Sitter metric with the constant $G$
proportional to the black hole mass and also the constant $F$
proportional to the cosmological constant, as expected.

Now let us see the behaviour of the scalar field at large
distances. It was observed in \cite{Henneaux:2004zi} that if the
scalar hair has a logarithmic form then its backreaction on the
metric changes the behaviour of the metric at large distances
having a slower asymptotic variation and in some cases even
changing the geometry of an AdS space.

The  asymptotic behaviour of the scalar field (\ref{scafiel}) is
given by
\begin{equation}
\phi \left( \rho \right) =\frac{1}{\sqrt{2}}\left( \frac{\nu }{\rho }-\frac{1%
}{24}\frac{\nu ^{3}}{\rho ^{3}}+O\left( \frac{1}{\rho ^{5}}\right)
\right)~.
\end{equation}
Then the metric (\ref{metricnew}) has quite different behaviour at
large distances. Calculating its  asymptotic $g_{\rho \rho }$
component  one finds
\begin{equation}
g_{\rho \rho }=\frac{l^{2}}{\rho ^{2}}+\frac{l^{4}}{\rho ^{4}}\left( -k-%
\frac{\nu ^{2}}{4l^{2}}\right) +O\left( \frac{1}{\rho
^{5}}\right)~.\label{expansion}
\end{equation}
We see that the geometry exhibits a deviation from AdS at large
distances in accordance with the observation in
\cite{Henneaux:2004zi}. However, the conserved charges are well
defined and finite  as we will see in next section, where we will
compute the mass and the entropy of our black hole solutions using
the Euclidean formalism. We will also study in the next section
the phase transitions between the asymptotically AdS
four-dimensional black holes  with scalar hair and the black hole
solutions without hair.

\section{Thermodynamics}
\subsection{Mass and entropy from boundary terms}

In order to calculate the mass and entropy of our black hole
solution using the Euclidean formalism, we will transform the
metric (\ref{metricnew}) to a more suitable form. Performing the
change of coordinates
\begin{equation}
\rho =\frac{\nu e^{\nu /(2r)}}{e^{\nu /r}-1}~,
\end{equation}
the metric (\ref{metricnew}) acquires the form
\begin{equation}\label{metric44}
ds^{2}=N\left( r\right) \left( -B\left( r\right)
dt^{2}+\frac{1}{B\left( r\right) }dr^{2}+r^{2}d\sigma ^{2}\right)
~,
\end{equation}
where
\begin{equation}
N\left( r\right) =\frac{\nu ^{2}e^{\nu /r}}{r^{2}\left( e^{\nu
/r}-1\right) ^{2}}~,
\end{equation}
\begin{equation}\label{Br}
B\left( r\right) =\frac{kr^{2}}{\nu ^{2}}\left( e^{\nu /r}+e^{-\nu
/r}-2\right) +Fr^{2}+\frac{2G}{\nu ^{2}}r-\frac{Gr^{2}}{\nu
^{3}}\left( e^{\nu /r}-e^{-\nu /r}\right)~ .
\end{equation}
Note that metric (\ref{metric44}) is symmetric under the change $\nu$ by $-\nu$, and the scalar field in the new coordinates becomes
\begin{equation}
\phi \left( r\right) =\frac{\nu }{\sqrt{2}r}~.
\end{equation}

We go to Euclidean time $t \rightarrow it$ and we consider the
action
\begin{equation}
I=\int \left( \pi ^{ij}\dot{g}_{ij}+p\dot{\phi}-NH-N^{i}H_{i}\right)
d^{3}xdt+B_{surf}~,
\end{equation}
where $\pi ^{ij}$ is the conjugate momenta of the metric and $p$
is the conjugate momenta of the field;  $B_{surf}$ is a surface
term. So, by considering the metric
\begin{equation}
ds^{2}=N^{2}\left( r\right) f^{2}\left( r\right) d\tau ^{2}+f^{-2}\left(
r\right) dr^{2}+R^{2}\left( r\right) d\sigma ^{2}~,
\end{equation}
where
\begin{eqnarray}
N\left( r\right) =\frac{\nu ^{2}e^{\nu /r}}{r^{2}\left( e^{\nu
/r}-1\right) ^{2}}~,\,\,
f^{2}\left( r\right) =\frac{r^{2}\left( e^{\nu /r}-1\right)
^{2}}{\nu ^{2}e^{\nu /r}}B\left( r\right)~,\,\,
R^{2}\left( r\right) =\frac{\nu ^{2}e^{\nu /r}}{\left( e^{\nu
/r}-1\right) ^{2}}~,
\end{eqnarray}
and with  a periodic $\beta=1/T$ where $T$ is the temperature, the
action becomes
\begin{equation}
I=-\beta \sigma \int_{r+}^{\infty }N\left( r\right) H\left( r\right) dr+B_{surf}~,
\end{equation}
where $\sigma $ is the area of the spatial 2-section. We now
compute the action when the field equations hold. The condition
that the geometries which are permitted should not have conical
singularities at the horizon imposes
\begin{equation}\label{T}
T=\frac{B^{\prime }\left( r_{+}\right) }{4\pi }~.
\end{equation}
So, by using the grand canonical ensemble we can fix  the
temperature. The variation of the surface term yields
\begin{equation}
\delta B_{surf}=\delta B_{\phi }+\delta B_{G}~,
\end{equation}
where
\begin{equation}
\delta B_{G}=\beta \sigma \left[ N\left( RR^{\prime }\delta f^{2}-\left(
f^{2}\right) ^{\prime }R\delta R\right) +2f^{2}R\left( N\delta R^{\prime
}-N^{\prime }\delta R\right) \right] _{r+}^{\infty }~,
\end{equation}
\begin{equation}
\delta B_{\phi }=\beta \sigma NR^{2}f^{2}\phi ^{\prime }\delta \phi~ .
\end{equation}
With an asymptotic behaviour of the metric function $f(r)$ given
by
\begin{equation}
f^{2}\left( r\right) =Fr^{2}+k+\frac{F\nu ^{2}}{12}-\frac{G}{3r}+O\left(
\frac{1}{r^{2}}\right)~,
\end{equation}
we find
\begin{equation}
\delta B_{G\infty }=\beta \sigma \left( \frac{F\nu r}{2}+O\left( \frac{1}{r}%
\right) \right) \delta \nu +\beta \sigma \left( -\frac{1}{3}+O\left( \frac{1%
}{r^{2}}\right) \right) \delta G~,
\end{equation}
and
\begin{equation}
\delta B_{\phi \infty }=\beta \sigma \left( -\frac{F\nu r}{2}+O\left( \frac{1%
}{r}\right) \right) \delta \nu ~.
\end{equation}
From the above expressions we deduce  the surface terms at
infinity
\begin{equation}
B_{surf \infty }=-\frac{\beta \sigma G}{3}~,
\end{equation}
and at the horizon
\begin{equation}
B_{surf r_+}=-\frac{\sigma }{4G_N}R^{2}\left( r_{+}\right)~.
\end{equation}
Therefore,  the Euclidean action reads
\begin{equation}
I=-\frac{\beta \sigma G}{3}+\frac{\sigma }{4G_N}R^{2}\left( r_{+}\right)~,
\end{equation}
and as the Euclidean action is related to the free energy through $I=-\beta F$, we obtain
\begin{equation}
I=S-\beta M~,
\end{equation}
where the mass $M$ is
\begin{equation}\label{M}
M=\frac{\sigma G}{3}~,
\end{equation}
and the entropy $S$ is
\begin{equation}\label{S}
S=\frac{\sigma }{4G_N}R^{2}\left( r_{+}\right)~.
\end{equation}
Notice that the mass (\ref{M}) is proportional to the constant $G$
as expected, and it is not related to the parameter $\nu$ of the
scalar field. This result is interesting compared to the results
obtained from the exact solutions of the action (\ref{action})
presented in Section 2. In all these solutions the mass and the
charge of the scalar field are related. This can be understood
from the fact that asymptotically the matter fields decouple from
the gravity sector so a kind of stealth mechanism is operating
\cite{AyonBeato:2004ig} allowing to found asymptotically AdS black
holes with primary hair.

\subsection{Phase transitions}

\begin{figure}[h]
\begin{center}
\includegraphics[width=0.35\textwidth]{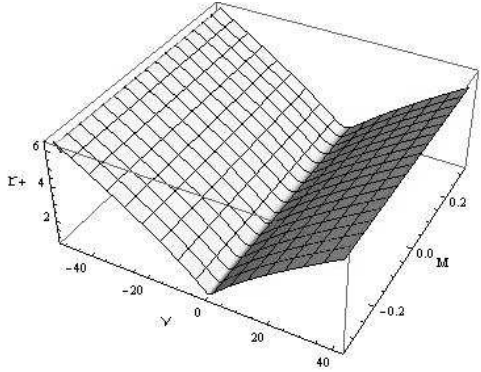}
\end{center}
\caption{The behaviour of $r_+$ respect to $\nu $ and to
the mass $M$ with $k=-1$, $l=1$, and $\sigma=1$.} \label{figura1}
\end{figure}

\begin{figure}[h]
\begin{center}
\includegraphics[width=0.35\textwidth]{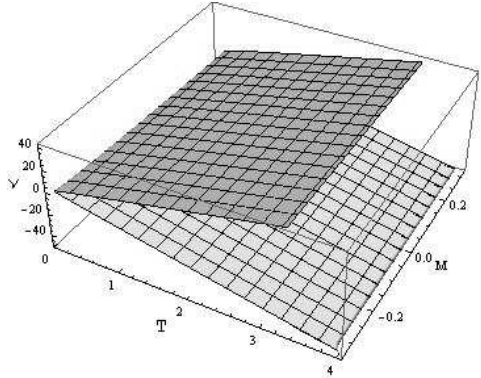}
\end{center}
\caption{The behaviour of $\nu $ as function of the temperature and
the mass $M$ with $k=-1$, $l=1$, and $\sigma=1$.} \label{figura2}
\end{figure}

Having the   temperature, mass and entropy for the asymptotically
AdS hairy black hole solutions   given  by Equations (\ref{T}),
(\ref{M}), and  (\ref{S}) respectively, we can study possible
phase transitions to known black hole solutions without hair. In
the absence of a scalar field the action (\ref{action}) for $k=-1$
has as a solution the topological AdS black hole
\cite{Mann:1996gj,Vanzo:1997gw,Mann:1997zn} with temperature,
entropy and mass  given respectively by
\begin{equation}
T=\frac{3}{4\pi l}\left( \frac{\rho _{+}}{l}-\frac{l}{3\rho _{+}}\right) ,%
\text{ \ }S_{TBH}=2\pi \sigma \rho _{+}^{2},\text{ \ }M_{TBH}=\sigma \rho
_{+}\left( \frac{\rho _{+}^{2}}{l^{2}}-1\right).
\end{equation}
So, the horizon radius $\rho _{+}=\frac{2\pi T}{3}+\sqrt{\left(
\frac{2\pi T}{3}\right) ^{2}+\frac{1}{3}}$ can be written as a
function of the temperature. To study the dependence of the
horizon radius of the hairy black hole with respect to the
parameter $\nu $ and to the mass $M$ of the hairy black hole, we
plot Fig. \ref{figura1}.
This Figure shows that the horizon radius takes the same value for
$\nu $ and $-\nu $, which is easily seen from Equation (\ref{Br}).
Also, we can see that there exist a minimum value of the horizon
when $\nu =0$ where the metric coincides with the topological
black hole. In Fig. \ref{figura2} we plot the parameter $\nu $ as
function of the temperature and the mass $M$ of the hairy black
hole. It is shown in this Figure that for a given mass and
temperature there are two values of the parameter $\nu $ for which
the temperature of the topological black hole is obtained.

\begin{figure}[h]
\begin{center}
\includegraphics[width=0.35\textwidth]{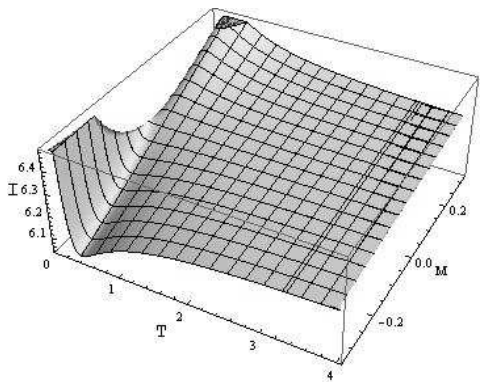}
\end{center}
\caption{The behaviour of Euclidean actions for the hairy black hole as function of the temperature and
the mass $M$ with $k=-1$, $l=1$, and $\sigma=1$.} \label{figura3}
\end{figure}

\begin{figure}[h]
\begin{center}
\includegraphics[width=0.35\textwidth]{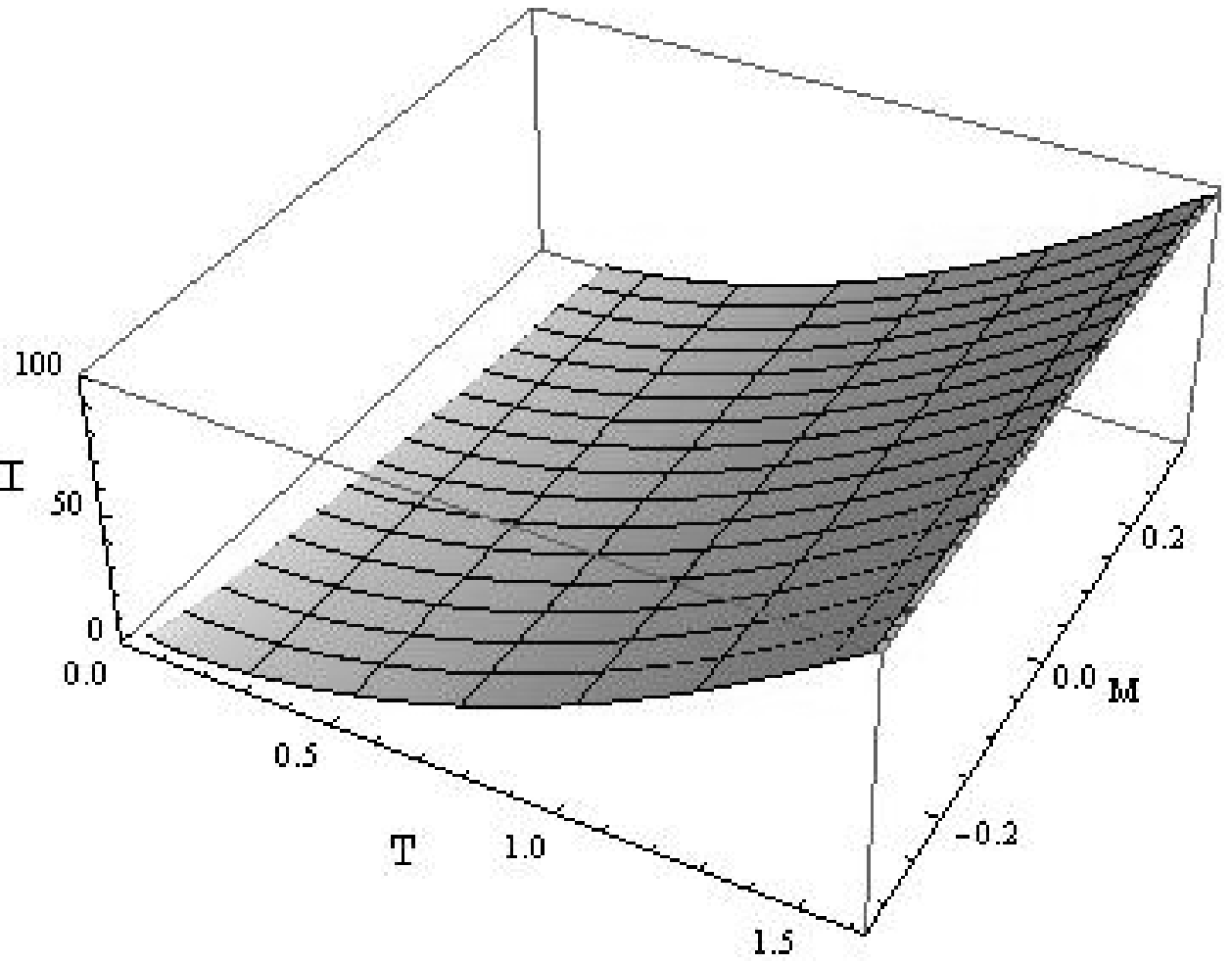}
\end{center}
\caption{The behaviour of Euclidean actions for the topological black hole as function of the temperature and
the mass $M$ with $k=-1$, $l=1$, and $\sigma=1$.} \label{figura4}
\end{figure}

\begin{figure}[h]
\begin{center}
\includegraphics[width=0.45\textwidth]{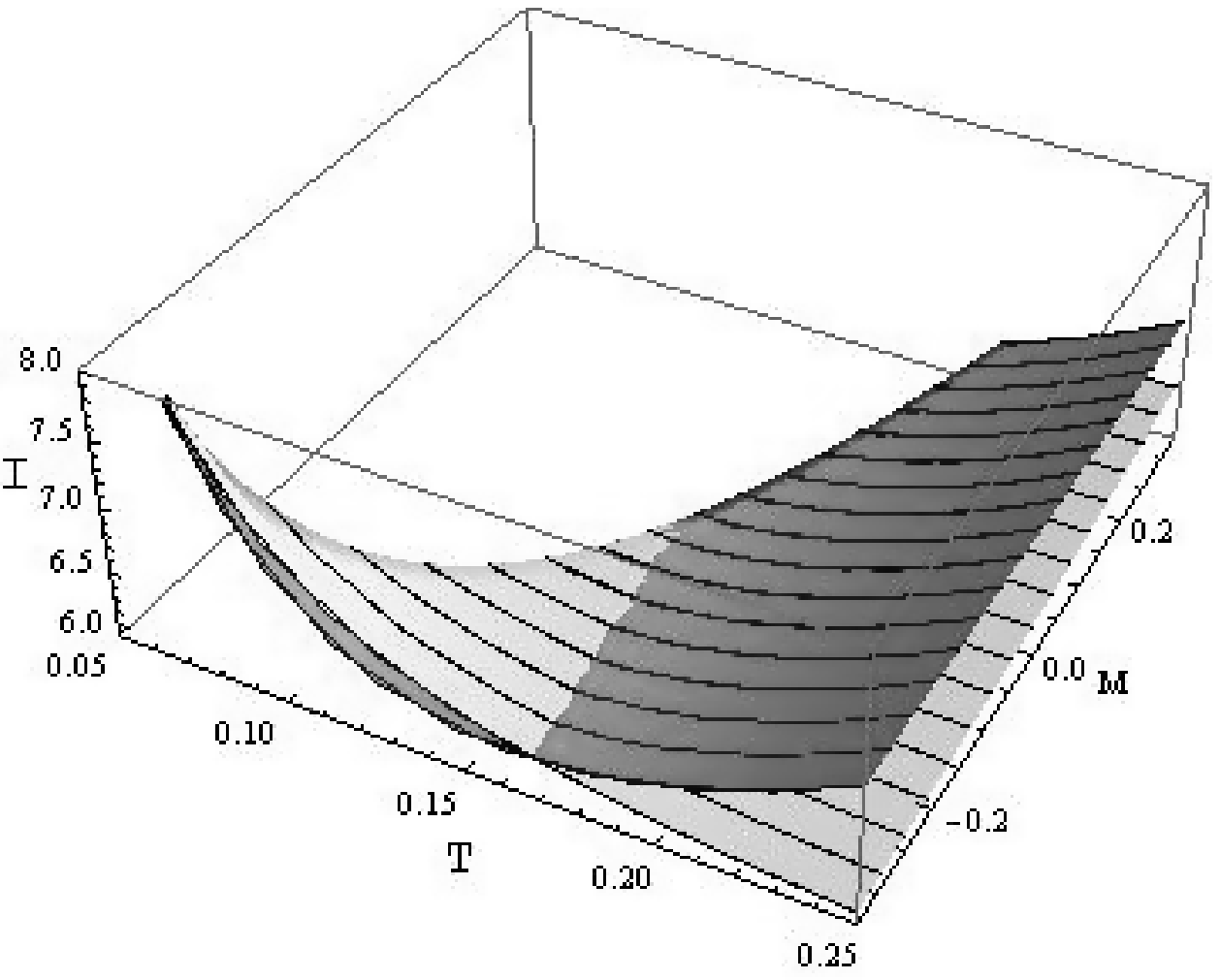}
\end{center}
\caption{The behaviour of Euclidean actions for the hairy black hole and the topological black hole as function of the temperature and
the mass $M$ with $k=-1$, $l=1$, and $\sigma=1$.} \label{figura5}
\end{figure}

The graphics for the Euclidean actions are showed in  Fig.
\ref{figura3} and \ref{figura4} for the black hole with scalar
hair and the topological black hole respectively and in   Fig.
\ref{figura5} we depict both actions in the same figure, in order
to see the range of values of black hole mass for which the phase
transitions exist.

Thus, from Fig. \ref{figura5}, we can see that there exists a
phase transition only for negative masses, and the hairy black
hole dominates for small temperatures, while for large
temperatures the topological black hole would be preferred. Also,
we can observe that the critical temperature at which this phase
transition takes place depends on the mass of the hairy black
hole. It is worth mentioning that the specific heat given by
\begin{equation}
C=-\frac{4\pi\nu^3e^{\frac{\nu}{r_+}}}{3(1+e^{\frac{\nu}{r_+}})(-2r_++2r_+e^{\frac{\nu}{r_+}}-\nu-\nu
e^{\frac{\nu}{r_+}})}
\end{equation}
is positive and therefore the asymptotically AdS
 black holes with scalar hair that we have found can always reach
thermal equilibrium with their surroundings and hence, are stable against thermal fluctuations.

Carrying out the same analysis for spherical horizons $k=1$ we
found that there is no phase transitions of the hairy
asymptotically AdS  black holes we found to Schwarzschild AdS
black hole. This result agrees with the findings in
\cite{Kolyvaris:2009pc} where only phase transitions of exact
hairy black hole solutions to black hole solutions with hyperbolic
horizons were found.

\section{Concluding comments}
\label{sec:conclution}

We have considered four-dimensional gravity theories where the
scalar field is minimally coupled to gravity along with a
self-interacting  potential. We have found a new family of
four-dimensional asymptotically AdS black holes with scalar hair.
These solutions asymptotically give the   Schwarzschild anti-de
Sitter solution. They characterized by a scalar field with a
logarithmic behaviour, being regular everywhere outside the event
horizon and null at spatial infinity,  and by a self-interacting
potential, which tends to the cosmological constant at spatial
infinity. Calculating the mass and entropy using the Euclidean
formalism we found that the mass of the black hole is not related
to the charge of the scalar field indicating that the scalar hair
is primary. Also, we have showed that there is a critical
temperature below which the gravitational system undergoes a phase
transition to a hairy black hole configuration, while above the
critical temperature an AdS black hole with hyperbolic horizon
dominates.

In this work we considered a particular profile for the scalar
field. The formalism developed  in Section 2 allows  us to
consider  other profiles for the scalar fields.  It would be
interesting to consider also exponential profiles for  the scalar
field which fall off at infinity quite fast  and study the
properties and the behaviour of possible asymptotically AdS black
holes with scalar hair.

The asymptotically AdS black hole solutions with scalar hair we
discussed in this work may help  to understand better the
generation of hairy black hole solutions and the mechanism of
scalar condensation which are necessary for the applications of
holography to condensed matter systems.  In most of these
applications, like the holographic superconductor, the probe limit
was considered in which the scalar field does not backreact on the
metric. The knowledge of exact or asymptotically  AdS  hairy black
hole solutions allows us to consider fully backreacted systems
\cite{Koutsoumbas:2009pa} which may give more information on the
boundary theory. It is also interesting to note\footnote{We thank
A. Anabalon for pointing out this behaviour of the potential to
us.} that the potential (\ref{m2potn}) can be embedded in
M-theory, as it arises from a consistent $U(1)^4$ truncation of
gauged $N=8$ supergravity \cite{Cvetic:1999xp}. This may help to
construct a realistic holographic superconductor using the
top-down approach.

\acknowledgments We would like to thank Theodoris Kolyvaris and
Minas Tsoukalas for valuable discussions and comments on the
draft. This work was funded by Comisi\'{o}n Nacional de Ciencias y Tecnolog\'{i}a through FONDECYT Grants 1110076 (JS, EP) and 11121148 (YV) and by DI-PUCV Grant 123713 (JS).


\end{document}